# Comment on "Evidence and Stability Field of fcc Superionic Water Ice Using Static Compression"

Alexander F. Goncharov [1] and Vitali B. Prakapenka [2]

[1] Earth and Planets Laboratory, Carnegie Science, Washington, DC 20015, USA.
[2] Center for Advanced Radiations Sources, University of Chicago, Chicago, Illinois 60637, USA.

Weck et al. [1] report on the existence and stability fields of two superionic (SI) phases of $H_2O$ ice at high P-T (P-T) conditions, which has been a topic of static and dynamic experiments and theoretical calculations (see Ref. [2] and references therein). They confirm Ref. [2] in that there are two SI phases with bcc and fcc oxygen sublattices with the stability at low- and high-P. However, they report on an extended stability field of fcc-SI ice toward lower T but no sign of it below 57 GPa. Here we argue that the reported phase boundaries of fcc-SI phase are not well experimentally justified due to difficulties to perform adequate X-ray diffraction (XRD) and radiometric measurements.

Weck et al. [1] used two different laser heating setups and sample loading techniques below and above 60 GPa. At 27-45 GPa their data agree with those of Ref. [2] concerning the stability field of bcc-SI (Fig. 1). However, unlike Ref. [2], these data at 40 and 45 GPa do not show any sign of fcc-SI phase (Figs. S3, S7 [1]) suggesting that the stability field of fcc-SI phase was not reached. Note, that at 45 GPa (Fig. S3 [1]) bcc-SI phase was reported at higher T (1652 K) than that where fluid phase was observed (1583 K) and no superionic phase is detected. These obvious inconsistencies demonstrate experimental difficulties to detect SI phases and control T in Ref. [1].

At 57 GPa, the data of Weck et al. are contradictory in the reported T range of bcc-SI and fcc-SI stability in two sets of data presented in Figs. 1, S6(top) compared to those of Fig. S4(top); the results of the second run agrees with Ref. [2] (Fig. 1). However, the first run shows an abrupt change in density of the fcc SI phase (Fig. 2) over a rather narrow T range (<150 K), which does not originate from the physical properties of the sample and sharply contrasts with the more gradual T dependencies of the second 57 GPa run, the 62 GPa run, and the results of Ref. [2]. Moreover, densities of the bcc-SI phases for the two 57 GPa runs differ substantially suggesting an abrupt drop in P during the first 57 GPa run (~10 GPa); such phenomena occur frequently in DAC laser heating experiments and such experiments should be discarded. At 62 GPa, the signatures of the fcc-SI phase are very subtle at 1653-2022 K [weak broad T independent *(111)* reflection only] making this observation unreliable. Note that the results of Refs. [2] and [1] agree concerning the fcc-SI phase stability region if the omitted in Figs. 4 and S7 of Ref. [1] higher T data at 57 GPa and 62 GPa are included (Fig. 1).

At 95 and 166 GPa, fcc-SI was observed at T lower than in Ref. [2]. This can be due to one-side heating/radiometry used in Ref. [1] (cf. [2]) because very large axial T gradients across the sample are present in a DAC cavity without thermal insulation. Indeed, the XRD signal of hot water phases is much weaker than that of cold bcc ice at 95 and 166 GPa, making problematic the reliable assignment of weak XRD peaks. At 95 GPa (Fig. S5 [1]), the Bragg peaks of LiF thermal insulation and the B-doped diamond coupler are very close to those of the purported fcc-SI. The observed peak splitting in XRD patterns at high T could be due to a large local thermal P (peaks of LiF shift to higher angles), chemical reaction of LiF with water, or even B diffusion. In a single measurements at 166 GPa where fcc phase was detected (Fig. 2 [1]), XRD shows a substantial increase in intensity of spurious Re gasket reflections at 2500 K, indicating a possible misalignment of the X-ray and laser heating spots. It is worth noting that unaccounted reflections were observed at high Ts in many high-T XRD patterns and images (Figs. 1, 2, S5 [1]) suggesting an ignored chemical contamination of the studied sample.

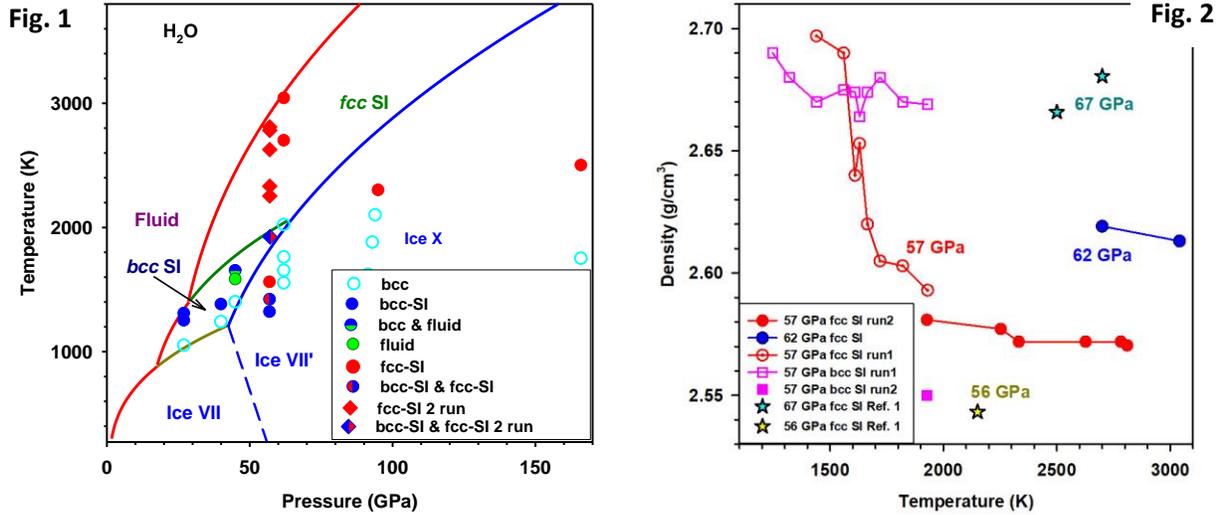

Fig. 1. Phase diagram of $H_2O$- the phase lines are from Ref [2]. Symbols are the P-T conditions of observations of superionic phases in Ref. [1].

Fig. 2. Density vs T plots of SI ices reported in Ref. [1] in comparison with the data of Ref. [2].